\documentclass{elsarticle}

\usepackage{epic}

\usepackage{latexsym}
\usepackage{amsmath}
\usepackage{epsfig}
\usepackage{amssymb}
\usepackage{enumerate}
\usepackage{natbib}
\usepackage[mathscr]{euscript}

\newtheorem{theorem}{Theorem}
\newtheorem{lemma}[theorem]{Lemma} 

\newtheorem{corollary}[theorem]{Corollary}
\newtheorem{proposition}{Proposition}

\newcommand{\PP}{{\mathbb P}}

\renewcommand{\Xi}{\mathcal{C}}

\begin{document}
\begin{frontmatter}

\title{Reflections on the extinction--explosion dichotomy}
\author{Mike Steel}
\address{Biomathematics Research Centre, University of Canterbury, Christchurch, New Zealand}

\date{\today}

\begin{abstract}
A wide range of stochastic processes that model the growth and decline of populations exhibit   a  curious dichotomy: with certainty either the population goes extinct or its size tends to infinity. There is a elegant  and classical theorem that explains why this dichotomy must hold under certain assumptions concerning the process.   In this note, I explore how these assumptions might be relaxed further in order to obtain the same, or a similar conclusion, and obtain both positive and negative results. \end{abstract}


\end{frontmatter}







\newpage
\section{Introduction}

The `merciless dichotomy' (Section 5.2 of \cite{had}) concerning extinction refers to a very general property of stochastic processes that describes the long-term fate of populations.  Roughly speaking, the result states that if there is always a strictly positive chance the population could become extinct in the future (depending, perhaps, on the current population size), then the population is guaranteed to either become extinct or to grow unboundedly large.  More precisely, a formal version of this result, due to Jagers (Theorem 2 of \cite{jag}), applies to any sequence $X_1, X_2, \ldots, X_n \ldots $ of non-negative real-valued random variables that are defined on some probability space and which is absorbing at 0 (i.e. $X_n=0 \Rightarrow X_{n+1}=0$ for all $n$).  It states that, provided: 
\begin{equation}
\label{strong}
\PP(\exists r: X_r=0|X_1, X_2, \ldots, X_n) \geq \delta_x>0 \mbox{ whenever $X_n \leq x$}
\end{equation} 
holds for all positive integers $n$,
then, with probability 1, either $X_n \rightarrow \infty$ or  a value of $n$ exists for which $X_k=0$ for all $k\geq n$ (notice that $\delta_x$ can tend towards 0 at any rate as $x$ grows). 
 This result  applies to a wide variety of stochastic processes studied in evolutionary and population biology (e.g. Yule  birth-death models, branching processes etc)  and the proof in \cite{jag} involves an elegant and short application of the  martingale convergence theorem. 

Note that the processes in  \cite{jag} (and here) need not be Markovian. Nevertheless, the lower-bound inequality condition in (\ref{strong}) has a Markovian-like
feature that it is required to hold for all values of $X_1, X_2, \ldots, X_{n-1}$ whenever $X_n$ is less than $x$. This raises the question of how much this uniform bounding across the previous history of the process might be relaxed without sacrificing the conclusion of certain extinction or explosion.   In this short note, we consider possible extensions of Jagers' theorem by weakening the assumption in (\ref{strong}).  Specifically, we will consider a  lower bound that conditions just on the event that $0<X_n \leq x$, either alone or alongside another variable that is dependent on (but less complete than) the past history $X_1, \ldots, X_{n-1}$.

First, we consider what happens if the probability in the  lower bound (\ref{strong}) were to condition just on $0<X_n \leq x$. In this case, we describe a positive result that delivers a slightly weaker conclusion than the original theorem of Jagers. We then show that the full conclusion cannot be obtained by lower bounds that condition solely on $0<X_n\leq x$ by exhibiting a specific counterexample.  
However, in the final section, we show that the full conclusion of Jagers' theorem can be obtained by conditioning on $0<X_n\leq x$, together with some partial information concerning the past
history of the process.

\section{A simple general lemma and its consequence for bounded populations}

We first present an elementary but general limit result, stated within the usual notation of  a probability space   $(\Omega, \Sigma, \PP)$ consisting of a sigma-algebra $\Sigma$ of `events'  (subsets of the sample space $\Omega$) and a probability measure $\PP$ (for background on probability theory, see \cite{borel}).
 
Suppose that $E_1, E_2,\ldots $ are {\em increasing} (i.e. $E_i \subseteq E_{i+1}$) and $E = \bigcup_{n=1}^{\infty}E_n$.  For example, suppose that $E_n$ is the event that some particular `situation' (e.g. extinction of the population) has arisen on or before a  given time step $n$ (e.g. day, year). These events are increasing and their union $E$ is the event that the `situation' eventually arises.
We are interested in when $\PP(E)=1$. A sufficient condition to guarantee this is to impose any non-zero lower bound on the  probability  that the `situation' arises at time step $n$ given that it has not done so already; in other words, to require that the conditional probability $\PP(E_n|\overline{E_{n-1}})$ is at least $\delta >0$ for all sufficiently large values of $n$ (throughout this paper an  overline denotes the complementary event).

On the other hand, it is equally easy to check that  if  $p_n=\PP(E_n|\overline{E_{n-1}})$ is allowed to converge to zero sufficiently quickly (so the probability of the `situation' first arising on day $n$ goes to zero sufficiently fast that $\sum_n p_n < \infty$),  then it is possible for  $\PP(E)<1$.   For example, if accidents occur independently and the probability of a particular accident is reduced each year by $1\%$ of its current value, then there is a positive probability that no accident will ever occur; but if the probability reduces at the rate $1, \frac{1}{2}, \frac{1}{3}, \frac{1}{4}, \frac{1}{5}, \cdots,$ then an accident is guaranteed to eventually occur (by the second Borel--Cantelli lemma).

Rather than placing some lower bound on the probability that the situation arises at time step $n$,
we can, following \cite{jag},  make a weaker assumption that if the situation has not happened yet,  there is always a non-vanishing chance that it will occur some time in the future 
(formally, requiring merely that $\PP(E|\overline{E_n})$ is uniformly bounded away from $0$).   For maximal generality, we also wish to avoid any
Markovian or independence assumptions.    The following lemma provides a sufficient condition for $\PP(E)=1$ without any further assumptions, and uses an elementary argument that will be useful later.

\begin{center}
\begin{figure}
\label{figure7}
\resizebox{7cm}{!}{
}
\end{figure}
\end{center}

\begin{lemma} 
\label{mike}
Suppose $E_n$ is an increasing sequence with limit $E$ and suppose that for some $\epsilon > 0$,  
$\PP(E|\overline{E_n}) \geq \epsilon$ holds for all $n \geq 1$. 
Then 
$\PP(E)=1$.
\end{lemma}

{\em Proof:}  Let $p_n = P(E_n)$.  Then, by the law of total probability:
\begin{center}$\PP(E) = \PP(E|\overline{E_n})(1 - p_n) + \PP(E|E_n)p_n$. \end{center}
Now, $\PP(E|E_n) = 1$ and, by assumption, $\PP(E|\overline{E_n}) \geq \epsilon$. Therefore:
\begin{center}$\PP(E) \geq \epsilon(1 - p_n) + p_n$.\end{center}

Since the events $E_n$ are increasing, a well known and elementary result in probability theory ensures that $\PP(E) = \lim_{n \to \infty} p_n$. So,  letting $n \to \infty$ in the previous inequality gives:
\begin{center}$\PP(E) \geq \epsilon (1 - \PP(E)) + \PP(E)$,\end{center}
which implies that $\PP(E) = 1$, as claimed. 
\hfill$\Box$

\subsection{Example 1}

Consider population of a species where $X_n$ denotes the size of the population at time step $n$.  The event $E_n = \{X_n=0\}$ is the event that the population is extinct
by time step $n$ and this  increasing sequence has the limit $E$ equal to the event of eventual extinction.  In this setting,   Lemma ~\ref{mike} provides the following special case
of Jagers' theorem. 

\begin{corollary}
\label{coro1}
Suppose that $X_1, X_2, \ldots, X_n$ is a sequence of non-negative real-valued random variables that are absorbing at 0 and are constrained to lie between $0$ and $M$.  Moreover, suppose that for some $\delta>0$ and all positive integers $n$ we have:
$\PP(\exists r: X_r=0|X_n  \neq 0) \geq \delta.$
Then, with probability 1, a value $n$ exists for which $X_k=0$ for all $k \geq n$.
\end{corollary}

\hfill$\Box$

\subsection{Remarks}

\begin{itemize}
\item[(a)]
One might view Lemma~\ref{mike} as a simple formulation of `Murphy's Law' -- the idea that if something bad can happen, it  will at some point (a popular claim often made in jest that has an interesting history \cite{murphy}). In  that context, $E_n$ is simply the event that the `bad thing' has happened on or before day $n$.

\item[(b)]
The proof of Proposition~\ref{mike} shows that
$\lim_{n \rightarrow \infty} \PP(E|\overline{E_n})>0 \Longrightarrow \PP(E)=1.$ The converse also holds, provided that $\PP(E_n)<1$ for all $n$;  indeed under that restriction, a sharper limit can be stated:
$\PP(E)=1 \Longrightarrow \lim_{n \rightarrow \infty} \PP(E|\overline{E_n})=1.$  With a view towards Borel--Cantelli type results,  note also that one can have: $\sum_{n \geq 1} \PP(E|\overline{E_n}) = \infty$ and  $\PP(E) <1$, if, for example, $\PP(E_n) = q-\frac{1}{n}$,
where $q<1$.

\item[(c)]
A general characterisation for when $\PP(E)=1$ is the following result from \cite{bruss}.
\begin{proposition}
\label{nice}
If $E_n$ is an increasing sequence of events with limit  $E$, then 
$\PP(E)=1$ if and only if either
 $\PP(E_1)=1$ or 
 $\PP(E_i|\overline{E_{i-1}})=1$ for some $i$, or  
$\sum_{i=1}^\infty \PP(E_{t_i}|\overline{E_{t_{i-1}}})= \infty$ for some  strictly increasing sequence $t_i$.
\end{proposition}

\end{itemize}

\section{A convergence in probability result for $X_n$}

We now consider what happens if the population size  is not bounded above by some maximal value $M$ as in Corollary~\ref{coro1}.
  In this case, by weakening the conditioning in Inequality (\ref{strong}) to just $X\in (0,m]$, one can still derive a result a result concerning convergence in probability (rather than almost sure convergence) of the population size to 0 or infinity, as we now show.

\begin{proposition}    
\label{jag2}
Suppose that $X_1, X_2, \ldots, X_n$ is a sequence of non-negative real-valued random variables that are absorbing at 0, and that for each  positive integer $m$, there is a value $\delta_m>0$ for which the following holds for all values of $n$: 
\begin{equation}
\label{boundful1}
\PP(\exists r: X_r=0|X_n  \in (0,m]) \geq \delta_m.
\end{equation}
Then, for every $m \geq 1$, we have
$\lim_{n \rightarrow \infty} \PP(X_n =0 \cup X_n >m) = 1.$
\end{proposition}
{\em Proof:}  Throughout this proof we will let $E$ denote the event $\{\exists r: X_r=0\}$.  The proof of Proposition~\ref{jag2} relies on the following result.

\begin{quote}{\bf Claim:}
Both $\PP(X_n=0|X_n \leq m)$ and $\PP(E|X_n \leq m)$  converge to 1 as $n \rightarrow \infty$.
\end{quote}

\noindent Proposition~\ref{jag2} follows directly from this claim, since, for any $m \geq 1$:
$$\PP(X_n =0 \cup X_n >m)  = \PP(X_n=0) + \PP(X_n > m)$$
$$\geq \PP(X_n=0|X_n \leq m)\PP(X_n\leq m)+ \PP(X_n >m).$$
By the claim, $ \PP(X_n=0|X_n \leq m)$ converges to 1 as $n$ grows, and so the previous inequality ensures that
$\lim_{n \rightarrow \infty} \PP(X_n =0 \cup X_n >m)  = 1,$
as required.  Thus it suffices to establish the claim.

\noindent{\em Proof of Claim:} 
Consider any  subsequence $n(k)$ of positive integers for which the bounded sequence $\PP(X_{n(k)} \leq m)$ has a limit. Such subsequences exist (by the Bolzano--Weierstrass theorem), and since $\lim \inf_{n \rightarrow \infty} \PP(X_n \leq m)>0$ (by (\ref{boundful1})) for all $m\geq 1$, the limit of $\PP(X_{n(k)} \leq m)$ for any such subsequence is strictly positive (this latter observation also ensures that some conditional probabilities below are well defined for large enough values of $k$). By the law of total probability:
$$\PP(E|X_{n(k)} \leq m) = \PP(E|X_{n(k)}=0) \PP(X_{n(k)}=0|X_{n(k)} \leq m) $$
$$+ \PP(E|X_{n(k)} \in (0, m]) \PP(X_{n(k)}>0|X_{n(k)} \leq m).$$
Thus if we let $p_k = \PP(X_{n(k)}=0|X_{n(k)} \leq m)$, then, by (\ref{boundful1}):
\begin{equation}
\label{pp1}
\PP(E|X_{n(k)} \leq m) \geq  1 \cdot p_k +\delta_m(1-p_k).
\end{equation}
Now, $p_{k} = \PP(X_{n(k)}=0)/\PP(X_{n(k)} \leq m)$ and so $\lim_{k \rightarrow \infty} p_{k} = \frac{\lim_{k\rightarrow \infty} \PP(X_{n(k)}=0)}{\lim_{k\rightarrow \infty} \PP(X_{n(k)} \leq m)}$,
since the numerator and denominator limits are non-zero. Moreover, we have $\PP(E)=\lim_{k\rightarrow \infty} \PP(X_{n(k)}=0)$ and so:
\begin{equation}
\label{pp2}
\lim_{k \rightarrow \infty} p_{k} = \lim_{k \rightarrow \infty} \PP(E|X_{n(k)} \leq m).
\end{equation}
Let $p$ denote the shared limit in Eqn. (\ref{pp2}). Then, from Inequality (\ref{pp1}) we have:
$$p \geq p + \delta_m(1-p),$$
which implies that $p=1$.  
Thus, for {\em all}  subsequences $n(k)$ of positive integers for which $\PP(X_{n(k)} \leq m)$ has a limit, this limit takes the same value (namely 1).  It follows from a well-known result
in analysis (e.g. Theorem 11, p. 67 of \cite{mal92})  that the full sequence $\PP(X_n \leq m)$ also converges to 1 as $n \rightarrow \infty$, and, therefore, so do the sequences   $\PP(X_n=0|X_n \leq m)$ and $\PP(E|X_n \leq m)$.
This establishes the two limit claims in the Claim, and so completes the proof of Proposition~\ref{jag2}.
\hfill$\Box$

Notice that Proposition~\ref{jag2} also implies Corollary~\ref{coro1} by taking $m = M$ and $\delta_m = \delta$ in (\ref{boundful1}).

\subsection{The conclusion of Proposition~\ref{jag2} cannot be strengthened to almost sure convergence.}

Suppose $X_1, X_2, \ldots, X_n \ldots$ is a sequence of non-negative real-valued random variables that satisfy the conditions described in Proposition~\ref{jag2}.  In this case, the proposition assures us that $X_n$ converges in probability either to 0 or to infinity.  This is a weaker conclusion than the statement that, with probability $1$, either $X_n=0$ for all sufficiently large $n$, or
 $X_n \rightarrow \infty$.  We now show, by an explicit example, that such a  stronger conclusion (which holds under the stronger condition (\ref{strong}) required for Jagers' theorem) need not hold under just the conditions described in Proposition~\ref{jag2}.   In other words, some additional conditioning on the past history of the process is required in order to secure the stronger conclusion (we describe this further in the next section).

\subsection{Example 2} 

Consider the following process.  Let $X_n^1, n \geq 1$ be a sequence of independent random variables with:
$$\PP(X^1_n) = \begin{cases}
1, & \mbox{ with probability } \frac{1}{n};\\
n, & \mbox{otherwise}.
\end{cases}
$$
For each $k \geq 2$, let
$X^k_n, n\geq 1$ be the (deterministic) random variables defined by:
$$\PP(X^k_n) = 
\begin{cases}
1, & \mbox{ with probability 1 for all $n  \in [1, \ldots, 2^k)$};\\
0, & \mbox{with probability 1 for all $n \geq 2^k$}.
\end{cases}
$$

Now, let $X_n$ be the stochastic process which selects $K=k$ with probability $\frac{1}{2^k}$ (for $k=1, 2,\ldots$) and then takes $X_n$ to be the process $X^K_n$ for all $n\geq 1$.

Firstly, note that this mixture process is well defined, since $\sum_{k \geq 1} \PP(K=k) = 1$. 
Next, observe that since $X^1_n = 1$ infinitely often (with probability 1) by the Borel--Cantelli Lemma (for independent random variables) and since there is a probability of $\frac{1}{2}$ that
$X_n = X^1_n$ for all $n$,  then, with probability $\frac{1}{2}$, $X_n$ does not converge to infinity or hit zero (note that $X^1_n \neq 0$ for any $n$, and $X^1_n$ returns to 1 infinitely often and
so does not tend to infinity).

Thus, to establish the claim regarding our example  it suffices to show that Inequality (\ref{boundful1}) applies.  This can be verified, and the details are provided in the Appendix.

\section{An extended extinction dichotomy theorem}
The example in the previous section  shows that  in (\ref{boundful1}) we need to supplement the condition $X_n \in ~(0, m]$ with some further information concerning the past history of the process, in order to guarantee eventual extinction or  $X_n \rightarrow \infty$. Here, we provide a mild extension of Theorem 2 of \cite{jag} by  conditioning on the number of times the process has dipped below each given value $m$ up to the present step of the process.  

\begin{theorem}
\label{jag3}
Suppose $X_1, X_2, \ldots, X_n \ldots$ is a sequence of non-negative real-valued random variables that are absorbing at 0.   For each positive integer $m \geq 1$,  let $\kappa_m(X_1, \ldots, X_{n-1})$ count the number of $X_1,X_2, \ldots, X_{n-1}$ that are less than or equal to $m$. Suppose that
for each positive integer $m$, there exists $\delta_m>0$ for which the following holds for all $n$.:
\begin{equation}
\label{boundful3}
 \PP(\exists r: X_r=0| X_n \in (0, m], \kappa_m(X_1, \ldots, X_{n-1})) \geq \delta_m.
\end{equation}
Then, with probability 1, either $X_n \rightarrow \infty$ or  a value of $n$ exists for which $X_k=0$ for all $k\geq n$.
\end{theorem}

{\em Proof:} 
For any strictly positive integers $n$ and $m$, let  $E_n$ be the event that $X_n=0$ and let $J_m$ be the event that $X_k \leq m$ for infinitely many values of $k$.
Notice that $E_n$ and $J_m$ are both  increasing  sequences.   Moreover, if we let $E= \bigcup_{n \geq 1}E_n$,  $J = \bigcup_{m \geq 1} J_m$  and $\overline{J} =\bigcap_{m \geq 1} \overline{J_m}$,  then
$E$ is the event that some $k$ exists such that $X_k=0$ and $\overline{J}$ is the event that $X_n \rightarrow \infty$.
We wish to show the following:
\begin{equation}
\label{ejeq}
\PP(E \cup \overline{J}) =1.
\end{equation}
Notice that: 
\begin{equation}
\label{ej}
E \subseteq J_m \mbox{ for each $m\geq 1$}.
\end{equation}  
Furthermore,
$\PP(E) >0$ by  Inequality (\ref{boundful3}) applied to $n=1$, and any value of $m\geq 1$ for which $\PP(X_1\leq m)>0$. Thus, from (\ref{ej}), 
$\PP(J_m)>0$ (and so $\PP(J)>0$ also),  so the conditional probabilities
$\PP(E|J)$ and $\PP(E|J_m)$ are well defined, and for each $m \geq 1$, the inclusion (\ref{ej}) gives:
\begin{equation}
\label{EJ2}
\PP(E) = \PP(E|J_m)\PP(J_m).
\end{equation}
We will show that:  
\begin{equation} 
\label{basic}
\PP(E|J_m) = 1 \mbox{ for each } m \geq 1,
\end{equation}
which, combined with  Eqn. (\ref{EJ2}), gives
$\PP(E) = \PP(J_m)$ for each $m \geq 1$.   Thus, since $\PP(J) = \lim_{m \rightarrow \infty} \PP(J_m)$ (recall $J_m$ are increasing),  we have
 $\PP(E) = \PP(J)$, and consequently
$\PP(E)+\PP(\overline{J})=1,$
since $E$ and $\overline{J}$ are mutually exclusive. In this way  we obtain the required identity (\ref{ejeq}) that establishes the theorem.

Thus it suffices to establish Eqn. (\ref{basic}). For this we employ a coupling-style argument. 
For each positive integer $m$,  we will associate to $X_n$ a second sequence of random variables $Y_k, k\geq 1$ as follows.
Let $O_m= \{n \geq 1:  X_n \leq m\}$, and for each  $k \leq |O_m|,$ let 
$Y_k = X_{\nu(k)}$ where the random variable $\nu(k)$ is the $k^{\rm th}$ element of $O_m$ under the  natural ordering of the positive integers.
  If $O_m$ is finite, then set $Y_k =0$ for all $k>|O_m|$ (notice that
this will not occur when we condition on $J_m$ below).

We may assume that the joint probability $\PP(Y_k \neq 0, J_m)$ is strictly positive; otherwise $\PP(Y_k=0|J_m)=1$ and so (\ref{basic}) holds, since 
$\PP(Y_k=0|J_m) \leq \PP(E|J_m)$.  Consequently, the conditional
probabilities are well defined in the following equation: 
\begin{equation}
\label{big}
\PP(E|Y_k \neq 0, J_m) = \sum_{n \geq 1} \PP(E| X_n \in (0, m], \nu(k)=n, J_m)\cdot \PP(\nu(k)=n|Y_k \neq 0, J_m).
\end{equation}
From (\ref{ej}) and (\ref{boundful3}), we obtain the following equality and inequality, respectively:
\begin{equation}
\label{one}
\PP(E| X_n \in (0, m], \nu(k) = n, J_m) \geq  \PP(E| X_n \in (0, m], \nu(k)=n) \geq  \delta_m,
\end{equation}
where the first inequality is from (\ref{ej}) and the second inequality is from (\ref{boundful3}), since conditioning on the conjunction $X_n \in (0, m], \nu(k) = n$ is equivalent to conditioning on the conjunction of $X_n \in (0,m]$ and
$\kappa_m(X_1, \ldots, X_{n-1}) = k-1$. 
 Substituting (\ref{one}) into the right-hand side of (\ref{big}) gives $\PP(E|Y_k \neq 0, J_m) \geq \delta_m$.
Thus, we have:
\begin{equation}
\label{good}
\PP(E|J_m) =\PP(E|Y_k \neq 0, J_m)\PP(Y_k \neq 0|J_m) + 1 \cdot \PP(Y_k=0|J_m)
\end{equation}
$$\geq \delta_m(1-p_k) + 1\cdot p_k,$$
where $p_k = \PP(Y_k=0|J_m),$ and where the factor $1$  is because, conditional on $J_m$, the event $E$ occurs whenever $Y_k=0$.
Now, $\{Y_k=0\}$ is an increasing  sequence in $k$, so if we let $\mathcal{Y}:= \bigcup_{k\geq 1}\{Y_k=0\}$,
then:
\begin{equation}
\label{helps1}
p:= \lim_{k \rightarrow \infty}  p_k=\PP({\mathcal Y}|J_m) .
\end{equation}
Moreover:
\begin{equation}
\label{helps2}
\PP(E|J_m) = \PP({\mathcal Y}|J_m).
\end{equation}
Applying (\ref{helps1}) and (\ref{helps2}) into (\ref{good})  gives:
$p \geq \delta_m(1-p)+p,$ which, in turn, implies that $p=1$ (since $\delta_m>0$).
Thus, $\PP(E|J_m)=p=1$, which  establishes (\ref{basic}) and so completes the proof.

\hfill$\Box$

\section{Concluding remarks}
Notice that Theorem~\ref{jag3} implies Theorem 2 of \cite{jag}, since the lower bound (\ref{boundful3}) involves conditioning on aggregates of values for  $X_1, \ldots, X_{n}$, so it holds automatically under the lower bound (\ref{strong}).  Notice also that the proof of Theorem~\ref{jag3}, though longer than the elegant martingale argument for Theorem 2 of \cite{jag}, requires merely elementary notions in probability.

It turns out that the collection of random variables $\kappa_m(X_1, \ldots, X_k)$ across all (real) values of $m$ and all integer values of $k$ between 1 and $n$ suffices to determine the sequence of random variables $X_1, \ldots, X_n$ (by induction on $k$), so it is not  immediately clear that Theorem~\ref{jag3} really allows greater generality than  Theorem 2 of \cite{jag}.  Therefore we provide an example to show that this is indeed the case.  Informally, the extra generality in Theorem~\ref{jag3}, arises from imposing fewer inequalities:  in (\ref{boundful3}) there are $n$ inequalities corresponding to the $n$ possible values that $\kappa_m(X_1, \ldots, X_{n-1})$ can take, while in (\ref{strong}), there are
potentially infinitely many, corresponding to all possible values for $X_1, \ldots, X_{n-1}$ (and for $X_n \leq x$).

\subsection{Example 3}

Roughly speaking, the stochastic process we will construct becomes extinct unless it oscillates regularly within a fixed range for an initial period, and the longer that it oscillates the greater the chance that it will escape to infinity rather than become extinct.  We show that such a process satisfies (\ref{boundful3}) but not (\ref{strong}). 

First,  consider a simple Markov chain $Y_n$ on the three states $0,1,2$ that starts in state 2 (i.e.  $Y_1=2$ with probability 1) and 
with transition probabilities described as follows:
\begin{itemize}
\item 0 is an absorbing state;
\item from state 1 or state 2, the next state is chosen with equal probability ($\frac{1}{3}$) from 0,1,2.
\end{itemize}
Thus, with probability 1,  a value $n$ exists for which $Y_k =0$ for all $k\geq n$.

We will say that  a sequence of values $y_1, y_2, y_3, \ldots, y_k$ from $\{1,2\}$ is a  {\em terminated flip sequence} (of length $k$) if $y_1 = 2$ and $y_{i} = y_{i-1}$ only for $i=k$.  For example $(2,1,2,1,2,1,2,1,1)$ and $(2,1,2,1,2,2)$ are terminated flip sequences of lengths nine and six respectively.

We use  $Y_n$  to define our process $X_n$ which takes non-negative integer values as follows.
If there is no value $N \geq 4$ for which $Y_1, Y_2, \ldots Y_N$ is a terminated flip sequence, then  set $X_n=Y_n$ for all $n$; in which case  $X_n$ absorbs at 0 with probability 1.
On the other hand, if a value $N \geq 4$ exists for which $Y_1, Y_2, \ldots Y_N$ is a terminated flip sequence, then, conditional on this value of $N$, 
$X_n = Y_n$ for all $n \leq N$, and for $n> N$,  $X_n = Z^N_{n-N}$, where $Z^N_1, Z^N_2, \ldots$ is a second Markov chain on the state space $\{0\} \cup \{ N-1, N, N+1, N+2,\ldots\}$.
This second chain has $Z_1= N-1$ (with probability 1), and has transitions from each state $i \geq N-1$ to $0$ and to $i+1$ with probabilities of   $2^{-i}$ and $1- 2^{-i}$, respectively.

Notice that, although the process $X_n$ is absorbing at $0$, it fails to satisfy (\ref{strong}) since, for an terminated flip sequence $(x_1,x_2, \ldots, x_n)$, of length 4 or more, 
we have $x_n \leq 2$ and yet:
$$\PP(\exists r: X_r=0| \wedge_{i=1}^n \{X_i =x_i\})  = \sum_{j=n-1}^\infty \frac{1}{2^j} \rightarrow 0, \mbox{ as } n \rightarrow \infty.$$

To show that $X_n$ satisfies (\ref{boundful3}), we consider the cases $m=1$, $m=2$ and $m>2$ separately.
For $m=1$,  (\ref{boundful3}) is equivalent to the following inequality  holding for all $n \geq 1$:
\begin{equation}
\label{m1}
 \PP(\exists r: X_r=0| X_n =1, \kappa_1(X_1, \ldots, X_{n-1})) \geq \delta_1>0.
 \end{equation}
Now, if $\kappa_1(X_1, \ldots, X_{n-1}) \neq \lfloor (n-1)/2\rfloor$ then $X_1, \ldots, X_n$ cannot be a terminated flip sequence, and so, with probability at least $\frac{1}{3}$, we have
$X_{n+1} = 0$.  On the other hand, if $\kappa_1(X_1, \ldots, X_{n-1}) = \lfloor (n-1)/2\rfloor$ then the probability that $Y_1, Y_2, \ldots, Y_n$ is a terminated flip sequence of length 4 or more is bounded away from 1 as $n$ grows, and so the event
$\{\exists r: X_r=0\}$ has a probability that is bounded away from $0$ for all $n$ when we condition on $\kappa_1(X_1, \ldots, X_{n-1})$ and $X_n = 1$.
Thus a value $\delta_1>0$ can be chosen to satisfy (\ref{m1}) for all $n \geq 1$.

For $m=2$,  (\ref{boundful3}) is equivalent to the following inequality  holding for all $n \geq 1$:
\begin{equation}
\label{m2}
\PP(\exists r: X_r=0| X_n  \in (0,2])\geq  \delta_2>0.
\end{equation}
Notice that $\kappa_2$ has vanished, since $X_n \in (0,2]$ implies that $\kappa_2(X_1, \ldots, X_{n-1}) = n-1$ with probability 1.
Now, conditional on $X_n \in (0,2]$, the probability that $Y_1, Y_2, \ldots, Y_n$ is a terminated flip sequence of length 4 or more is bounded away from 1 as $n$ grows, and so the event
$\{\exists r: X_r=0\}$ has a probability that is bounded away from $0$ for all $n$ when we condition on$X_n \in \{1,2\}$. Thus a value $\delta_2>0$ can be chosen 
to satisfy (\ref{m2}) for all $n \geq 1$.

Finally, for each $m>2$, for all $n \geq 1$:
$$\PP(\exists r: X_r=0| X_n  \in (0,m], \kappa_m(X_1, \ldots, X_{n-1})) \geq 2^{-m}>0,$$
so we can set $\delta_m = 2^{-m}$ for all $m>2$. 
In summary, for all values of $m$,  $X_n$ satisfies (\ref{boundful3}) for all $n$, as claimed.

\section{Acknowledgments}
I thank Elchanan Mossel for several helpful comments concerning an earlier version of this manuscript, and Elliott Sober for some motivating discussion. I also thank the Allan Wilson Centre for funding support for this work.

\section{Appendix: Proof that Example 2 satisfies Inequality (\ref{boundful1})}

 Firstly, if $m<1$, then conditioning on $X_n\leq m$ is equivalent to conditioning on $K>1$ and so we can take  any positive value  for $\delta_m$ (even  $= 1$) and satisfy Inequality (\ref{boundful1}).
 
Next, suppose that $m \geq 1$, and, for any $n \geq 1$, write:
\begin{equation}
\label{eq0}
n = 2^q +r, \mbox{ where } 0\leq r < 2^q, q\geq 0.
\end{equation}
SInce $\PP(\exists r: X_r=0|X_n  \in (0,m]) = \PP(K>1|X_n\in (0,m])$ we have:
\begin{equation}
\label{eq1}
\PP(\exists r: X_r=0|X_n  \in (0,m]) =  \sum_{k \geq 2}  \PP(K=k|X_n\in (0,m]),
\end{equation}
and, from Bayes' identity:
\begin{equation}
\label{eq2}
\PP(K=k|X_n\in (0,m]) = \frac{\PP(X_n \in (0,m]|K=k)\PP(K=k)}{\PP(X_n \in (0,m])}.
\end{equation}

Now, for any $k \geq 2$ (and still with $m>1$): 
\begin{equation}
\label{eqx}
\PP(X_n \in (0,m]|K=k) =
 \begin{cases}
1, & \mbox{ provided $k \geq q+1$};\\
0, & \mbox{otherwise};
\end{cases}
\end{equation}
since $X_n^k=1$ with probability 1 for all $n  \in [1, \ldots, 2^k)$.
Consequently, the numerator of (\ref{eq2}) equals  $\frac{1}{2^k}$ ( $=\PP(K=k)$)  when $k \geq q+1$ and is zero otherwise.

Now, the denominator of (\ref{eq2}), namely $\PP(X_n \in (0,m])$, can be written as:
\begin{equation}
\label{eq3}
 \left[\sum_{k \geq 2} \PP(X_n \in (0,m]|K=k)\PP(K=k)\right] + \PP(X_n\in (0,m]|K=1)\PP(K=1).
\end{equation}
From (\ref{eqx}), the first term in (\ref{eq3}) is:
\begin{equation}
\label{eq4}
\sum_{k \geq 2} \PP(X_n \in (0,m]|K=k)\PP(K=k)= \sum_{k \geq q+1} \frac{1}{2^k} = \frac{1}{2^q}.
\end{equation}
Regarding the second term in  (\ref{eq3}), observe that:
$$\PP(X_n\in (0,m]|K=1) = 
 \begin{cases}
\frac{1}{n}, & \mbox{ provided $n > m$};\\
1 , & \mbox{if $n \leq m$. }
\end{cases}
$$
Therefore, recalling (\ref{eq0}), the second term in (\ref{eq3}) is:
\begin{equation}
\label{eq5}
 \begin{cases}
\frac{1}{n} \times \frac{1}{2} = \frac{1}{2^{q+1}+2r} \leq  \frac{1}{2^{q+1}} , & \mbox{ provided $n >m$};\\
1 \times \frac{1}{2}, & \mbox{when $n \leq m$}.
\end{cases}
\end{equation}
 Consequently, by combining (\ref{eq3}), (\ref{eq4}) and (\ref{eq5})  into  (\ref{eq2}) (and noting again that $\sum_{k \geq q+1} \frac{1}{2^k} = \frac{1}{2^q}$) we have that if $n >m$, then
$\sum_{k \geq 2}  \PP(K=k|X_n\in (0,m]) \geq 
\frac{\frac{1}{2^q}}{\frac{1}{2^q} + \frac{1}{2^{q+1}}} \geq \frac{1}{2},$
while if $n \leq m$, then 
$\sum_{k \geq 2}  \PP(K=k|X_n\in (0,m]) \geq \frac{\frac{1}{2^q}}{\frac{1}{2^q} + \frac{1}{2}} \geq  \frac{\frac{1}{2^q}}{1+ \frac{1}{2}} \geq \frac{2}{3}\cdot \frac{1}{2^q} \geq  \frac{2}{3m},$
where the last inequality is from $ m \geq n \geq 2^q$.
Thus, if we take $\delta_1 = \frac{1}{2}$ and $\delta_m = \frac{2}{3m}$ for each $m \geq 2$, then, 
from (\ref{eq1}), 
$ \PP(\exists r: X_r=0|X_n  \in (0,m])\geq \delta_m$
 for all $n, m$, 
 as claimed.

\end{document}